\documentclass[runningheads]{llncs}
\usepackage{graphicx}
\usepackage{amsmath}
\usepackage{algorithm}
\usepackage{algorithmic}

\usepackage{multirow}
\usepackage{tabularx}
\usepackage{tablefootnote}
\usepackage{enumitem}
\usepackage{amsmath,amssymb,amsfonts}

\begin{document}
\title{Explainable AI for Suicide Risk Assessment \\Using Eye Activities and Head Gestures}
%
%
\author{Siyu Liu\inst{1} \and
Catherine Lu\inst{1} \and
Sharifa Alghowinem\inst{2}\orcidID{0000-0002-9391-0163}\and
Lea Gotoh\inst{3} \and
Cynthia Breazeal\inst{2}\orcidID{0000-0002-0587-2065} \and
Hae Won Park\inst{2}\orcidID{0000-0001-9638-1722}}
\authorrunning{Liu et al.}
%
\institute{Massachusetts Institute of Technology, Cambridge MA 02139, USA \email{eliu24@mit.edu, czlu@mit.edu} \and
Personal Robotics Group, MIT Media Lab, Cambridge MA 02139, USA \email{sharifah@media.mit.edu, cynthiab@media.mit.edu, haewon@media.mit.edu} \and
Japan Broadcasting Corporation, Tokyo, Japan\\
\email{gotou.r-km@nhk.or.jp} 
}
\maketitle              
\begin{abstract}

The prevalence of suicide has been on the rise since the 20th century, causing severe emotional damage to individuals, families, and communities alike.
Despite the severity of this suicide epidemic, there is so far no reliable and systematic way to assess suicide intent of a given individual. 
Through efforts to automate and systematize diagnosis of mental illnesses over the past few years, verbal and acoustic behaviors have received increasing attention as biomarkers, but little has been done to study eyelids, gaze, and head pose in evaluating suicide risk. This study explores statistical analysis, feature selection, and machine learning classification as means of suicide risk evaluation and nonverbal behavioral interpretation.
Applying these methods to the eye and head signals extracted from our unique dataset, this study finds that high-risk suicidal individuals experience psycho-motor retardation and symptoms of anxiety and depression, characterized by eye contact avoidance, slower blinks and a downward eye gaze. By comparing results from different methods of classification, we determined that these features are highly capable of automatically classifying different levels of suicide risk consistently and with high accuracy, above 98\%. Our conclusion corroborates psychological studies, and shows great potential of a systematic approach in suicide risk evaluation that is adoptable by both healthcare providers and naïve observers.

\keywords{Affective Computing \and Suicide Risk \and Nonverbal Behaviour \and Explainable AI.}
\end{abstract}
\section{Introduction}

Among the many suicide attempts each year, 800,000 people die \cite{WHOsuicide}. Often, they suffered from depression, anxiety, or have histories of self-injury for years \cite{guan2012nonsuicidal}. Those that pass leave tremendous emotional strain to their families and communities. 
Japan is at the center of this suicide epidemic. Historical trauma, war, unemployment, natural disasters, high stress, all contribute to the alarmingly high rate of 20,000 deaths per year \cite{Matsubayashi2020}.
%
Yet, suicide is preventable. Given the association between suicide inclination and mental disorder, early identification and intervention can mitigate many tragedies \cite{160658}. 
%

Current suicide assessment criteria --- including Behavioral Health Screening (BHS), Manchester self harm rule (MSHR), and Södersjukhuset self-harm rule (SOS-4) --- lack specificity and sensitivity, and are difficult to administrate, and those with suicide inclinations are often blamed for not seeking help \cite{doctorsinterview,e0180292}. 
This hinders early detection and intervention.


Recent efforts to automate suicide risk assessment using machine learning on metadata has shown promising results, yielding high classification accuracy ($>$ 90\%) in suicide ideation prediction using electronic medical records and co-morbidity biomarkers \cite{PMC7460360}. 
However, significant challenges exist in application of these results, as no model has successfully been transferred into new settings or populations \cite{PMC7460360,fonseka2019utility}. While recent works have explored using objective verbal and acoustic markers for suicide risk assessment \cite{huang2007hunting,chowdhury2007trec,gomez2014language}, to our best knowledge, the three studies led by Laksana \cite{7961819}, Eigbe \cite{eigbe8373878} and  Shah \cite{shah2019multimodal} are the only works that investigate observable facial behavior markers. 


The purpose of this study is to construct a systematic and interpretable approach for suicide risk assessment. 
We investigate whether nonverbal behaviors from the eyelids, head pose, and eye gaze hold discriminative power for evaluating suicidal risks, considering a wider array of behavioral features than previous works using statistical analysis, feature selection, and classification with machine learning.
Given the interpretability approach of this work, we hope to propose a systematic approach in suicide risk evaluation that is executable for both healthcare providers and naïve observers.
The main contributions of this research are as follows:
\begin{itemize}
    \item extract objective and interpretable behaviors that are discernible to naïve observers
    \item provide analyses using statistical methods, feature selection, and machine learning with a focus on explainability 
\end{itemize}

\section{Background}

Nonverbal indicators of suicide ideation have received far less attention than verbal ones. Laksana et al. found simple facial features including smiling and head and eyebrow movement effective for suicide risk assessment \cite{7961819}, and that suicidal individuals produce more non-Duchenne smiles. Eigbe et al. verified this result and also found that suicidal subjects look down more often than their non-suicidal counterparts \cite{eigbe8373878}.

Suicidal intent have strong correlation with other mental disorders, including depression and anxiety. 
Symptoms of anxiety and depression ---  high stress, fatigue, and social withdrawal --- have been observed among suicidal individuals as well \cite{doctorsinterview}. Particularly, statistical analysis and machine learning classification of the movement of facial and ocular landmarks found that 
depressed patients have narrower eye openings, longer duration of blinks, and slower and less frequent head movements that signify fatigue or eye contact avoidance  \cite{6738869,waxer1977nonverbal,alghowinem2013head}. Fossi et al. attribute this phenomenon as a form of psycho-motor retardation \cite{fossi1984ethological}. Additionally, by analyzing data collected with eye trackers, infra-red corneal reflection techniques, or human observation, many studies have confirmed that patients with anxiety disorder avoid eye contact and direct gaze at the person's face in a photo, but scan non-facial features and the surrounding environment more extensively \cite{MOUKHEIBER2010147,HORLEY200333,JUN2013193}. As depression and anxiety disorder are significant risk factors for suicidal thoughts and behaviors, we may leverage from nonverbal indicators of depression and anxiety to formulate our research.


To classify such indicators, machine learning has been utilized extensively. Automatic assessment of mental health disorders such as depression, schizophrenia, and bipolar disorder has primarily used machine learning classification to provide objective diagnosis \cite{low_bentley_ghosh_2020}. These efforts have been successful. For instance, Cohn et al. \cite{cohn_depression}, Abaei and Osman \cite{abaei_osman_bp}, and Shah et al. \cite{shah2019multimodal} classified for depression, bipolar disorder, and suicidal intent by analyzing visual, acoustic, and verbal features using machine learning techniques including support vector machines (SVMs), logistic regression, and convolutional neural-nets (CNNs). Given the correlation between these disorders and suicide, classification of verbal and nonverbal behaviors may yield good results for predicting suicidal risk.

Yet, few studies have attempted automatic machine learning classification of suicide risk. Laksana et al. \cite{7961819} and Eigbe et al. \cite{eigbe8373878} have achieved 40\% and 69\% accuracy in classifying suicidal, mentally ill, and control groups based on facial behaviors. In this paper, we explore combinations of different discriminative models (multi-layer perceptron and SVMs) and data processing methods to determine the best-performing classifier of suicide risk. We build upon the previous research by extracting an extensive array of nonverbal behaviors from head and eye movements. Then, we twice execute classification -- once with all  features and once with just 10\% selected through feature selection and statistical analysis. Specifically, we chose 10\% in order to only select the most prominent features. If the classification results obtained in both trials are similar, we deem the selected features to be representative of the task and capable of explaining the model.

\section{Methodology}

\begin{figure*}[t]
    \centering
    \includegraphics[width=0.99\textwidth]{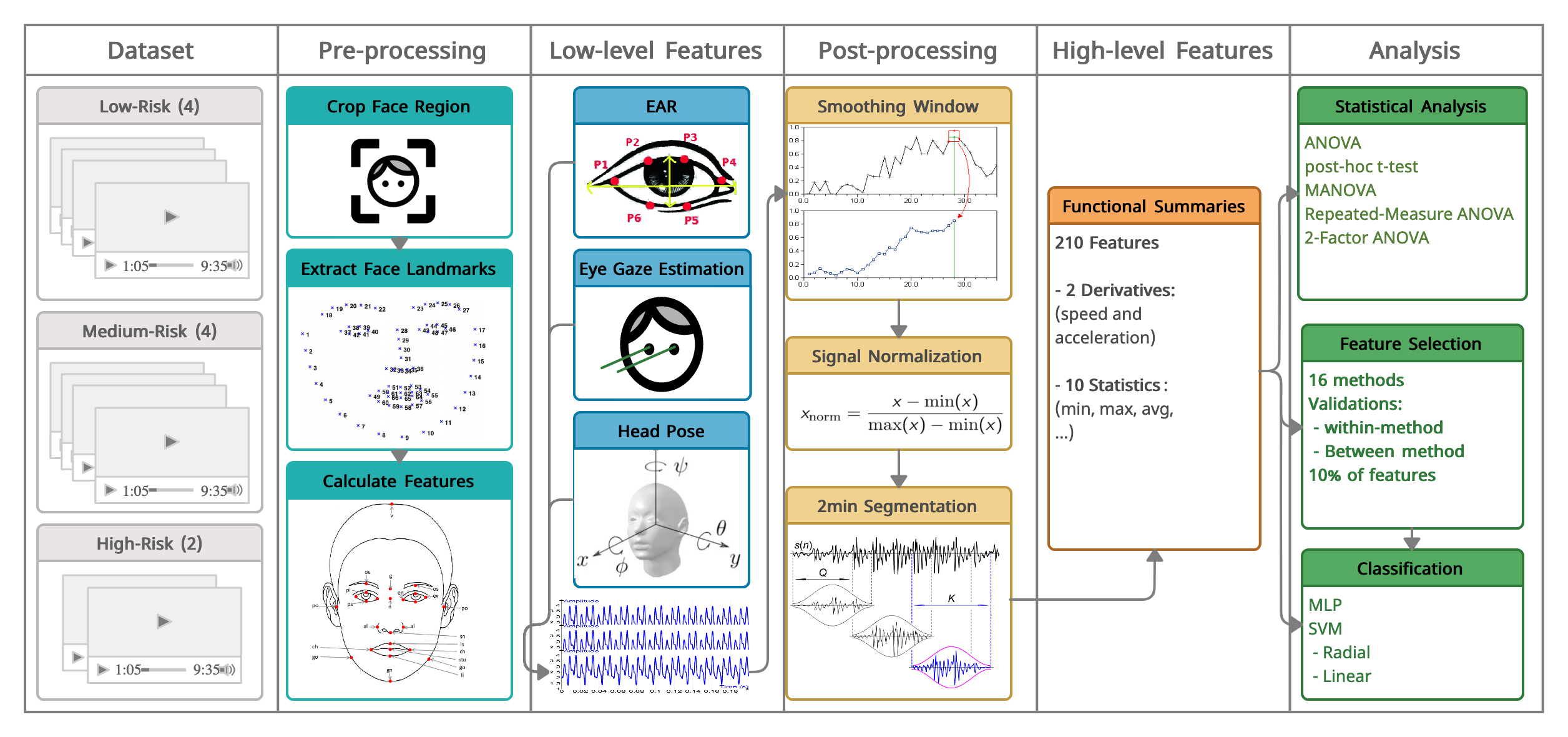}
    \caption{The flowchart shows the general steps and their sub-steps start from dataset acquisition, pre-processing of signals (cropping image, facial landmark extraction, and basic feature calculation), low-level feature extraction (EAR, eye gaze, head pose estimation), post-processing of signals (window smoothing, signal normalization, thin-slicing), high-level feature extraction through functional summaries, and ending with analysis and interpretation of results through statistical analysis, features selection, and classification.}
    \label{fig:methodology_process}
\end{figure*}

The overall process including pre-processing, low-level features extraction, post-processing and high-level feature summaries, and analysis is summarized in Fig.\ \ref{fig:methodology_process} and described in the following subsections.

\begin{table}[t]
\scriptsize
\centering
\caption{Number of Subjects, Interview Duration and Total Segments in each Suicide Risk Level in the Sub-Dataset used in this Work}
\begin{tabular}{@{}c | r |l |l @{}}
\textbf{Risk Level} &  \textbf{\# Subj} & \textbf{Interview Duration (min)}& \textbf{\# of 2 min Segments}\\
\hline
Low  & 4 & 60.80   & 54  \\
Medium & 4 & 88.00 & 63  \\
High & 2 & 31.16  & 27 \\
\hline
Total & 10 & 179.96  & 144 \\
\hline
\end{tabular}
\label{tab:dataset}
\end{table}

\subsection{Suicide Dataset Collection}\label{AA}
The dataset was collected in collaboration with Japan Broadcasting Corporation (NHK), 
who conducted interviews with Japanese participants aged 15 to 25 who are suffering from suicidal thoughts as part of their suicide awareness project. The project was advertised through social media, which invited people to share stories of their (or their loved ones') suffering on the project's website ``Face to Suicide.'' \footnote{http://www6.nhk.or.jp/heart-net/mukiau}. People who were willing to be further interviewed were recruited for the project. All participants provided consent via email communication. 
The selected participants were pre-interviewed (without camera recording) by two project directors for screening and suicide risk assessment. During the pre-interview, the participants were assessed using the “Suicidal Risk” subsection of the M.I.N.I. (Mini-International Neuropsychiatric Interview) \cite{lecrubier1997mini}, which consists of 6 questions about their negative thoughts (e.g., suicidal ideation and intensity). Upon consultation with psychiatrists, another 5 questions were added to the pre-interview about their history of self-harm and clinical visits, environment (e.g., age, family, socioeconomic status), and positive thoughts (e.g., trusted persons, hobbies, coping mechanisms). The two directors then categorized each participant as a low, medium, or high suicide risk individual. These assessments and potential conflicts were subsequently cross-validated and were resolved via discussion that accompanied psychiatrist consultation. 
A total of 14 participants were interviewed, once each; 4 of them requested to have their faces blurred for the recording. This work uses the 10 interviews with faces not blurred to study participants' nonverbal behaviors. Four participants were assessed as low risk, four at medium risk, and two at high risk (see Table \ref{tab:dataset}). The interviews were conducted in several locations with different room layouts (e.g., window and door location, furniture), but the relative position of the interviewer-interviewee were consistent --- both sat in chairs where they faced each other.

\subsection{Feature Extraction}

\paragraph*{\textbf{Pre-processing of feature extraction}}
High resolution videos were manually cropped to contain only the face region of the subject. For each frame of the processed video, we used single-shot face detection to locate a total of 68 facial landmarks, 
as proposed in \cite{zhang2017s3fd}. Frames where the face is not fully captured were skipped.
Additionally, using the facial landmarks, cropped images containing each eye were obtained from each valid frame, which served as inputs to calculate gaze direction (yaw and roll).

\paragraph*{\textbf{Low Level Feature Extraction}}
For each frame, Eye Aspect Ratio (EAR) and gaze direction were extracted for each eye, and head direction was extracted for the entire face image, as described below.
Proposed by Soukupova and Cech, EAR characterize the openness of an eye as a scalar quantity in an image\cite{cech2016real}, where this ratio is normalized against distance of the eye to the camera and eye size differences between the subjects. A larger EAR value indicates a wider opening. A similar approach was taken by \cite{6738869} for the analysis of eye movement for depression detection. Using the 6 eye-region landmark coordinates, EAR is computed based on the formula as follows, where $p_i$ are the 2D landmark locations as depicted in Fig.\ \ref{fig:methodology_process} (EAR).
\begin{equation}
\dfrac{\left\lVert p_2 - p_6\right\rVert - \left\lVert p_3 - p_5\right\rVert}{2\left\lVert p_1-p_4\right\rVert}
\label{eq1}
\end{equation}


To describe each eye's angle of gaze, two scalar quantities, pitch and yaw, were extracted per frame. We used \cite{zhang15_cvpr}'s multimodal CNN to estimate the eye gaze. The inputs of this CNN are the facial landmarks, normalized head angle vectors, and cropped eye images. 
This been trained on the MPIIGaze dataset that contains 213,659 images with diversity in appearance and illumination collected from everyday laptop usage of 15 participants~\cite{zhang2017mpiigaze}. 

Head pose was extracted in terms of three angle vectors --- pitch, yaw, and roll for each frame --- to represent the direction the subject is facing. Our study employed the technique outlined by \cite{Sugano3DGaze} to estimate the 3D head pose using 6 facial landmarks --- four eye corners and two mouth corners. By fitting generic 3D facial shape models onto our facial landmarks, we obtained angles of rotation of the head. Lastly, a distance to the face was calculated from the camera's focal point, based on the apparent size of the face.

\paragraph*{\textbf{Post-processing of the Signals}}
Low-level feature signals were post-processed to eliminate outliers and noise. For each low-level signal, a moving average was computed with a window size of 7 frames (window size 3 -- 20 were attempted). The moving average window was chosen empirically to reduce noise without eliminating actual movements. For features that are applicable to both eyes (EAR, eye pitch, and eye yaw), results obtained from the left and right eye were averaged to reduce the variability and noise of the two signals. This final 7 low-level features (avg. EAR, avg. eye pitch, avg. eye yaw, head distance, head pitch, head yaw, and head roll) were used in further analysis. In order to eliminate inter-subject variability including the participant's personality, mannerism, and appearance, we linearly scaled each signal from 0 to 1. 

\paragraph*{\textbf{High Level Feature Extraction}}
We segmented our data into small windows following the physiological thin-slicing theory \cite{ambady1992thin}, where a brief observation of a behavior (a thin slice) can be indicative to the physiological outcome at levels similar to the full observation. This process also serves as a mitigation for the small number of subjects in our dataset. As Shah et al.'s experiment has shown, 2-minute slices are just as effective as 5-minute slices in capturing features of the video~\cite{shah2019multimodal}. Hence, we segmented each subject's signal into 2-minute slices, with a one-minute overlap between two adjacent slices. The total number of segments per risk group is listed in Table \ref{tab:dataset}. A total of 210 statistical features, i.e., ``functionals'', were extracted to summarize the segments.
For each of the 7 post-process low-level features, 30 statistics were calculated, which are:

\begin{itemize}
    \item The speed and acceleration at which the signal moves. This is computed by taking the first and second derivatives with regard to the frame index.
    \item Maximum, minimum, range, mean, variance, standard deviation, skewness, kurtosis, number of peaks and valleys for the original signal and its derivatives.
\end{itemize}

These functionals capture the frequency and velocity of the movement of each signal, as well as the duration of continuous movement and their direction. 
The compilations of statistical features from 2-minute segments across videos from the same categories --- a statistical summary --- were used as samples for our classification, feature selection, and statistical analysis. One study employing a similar approach of statistical hypothesis testing and multimodal predictive modeling yielded promising results in predicting suicide intent using sliced videos from social media \cite{shah2019multimodal}.

\subsection{Feature Selection}\label{sec:FS}
A \textbf{feature selection framework (FS)} to systematically select the most representative features that are correlated to an independent variable, i.e. nonverbal behavior with suicide risks, was proposed in \cite{9253541}, which we replicated in this work.   While extracting a large number of features as an exploration of behaviors is a common practice in the AI community, it is not commonly practiced in the psychology field for diagnoses without confirming an assumption or a hypothesis. Such confirmation is done through statistical analysis that accounts for multi-test correction (e.g., Bonferroni correction), which might not be ideal for large feature spaces and small sample sizes. Therefore, the FS framework aimed to fill the gap by systematically aggregating the results from several statistical analyses with several methods of feature selection. The framework also serves as an interpretation tool through narrowing the feature space to the most meaningful features for a variable/class, not only by analyzing the features independently, but also analyzing the relationship between features (e.g., removing redundant features, finding a combination of features that correlate together, etc.). Given the two-step validation approach, i.e., within and between methods validation, the sensitivity of any method to the sample size is mitigated. That is, when a feature is not stable enough for the selection process, the framework excludes it.
 
In this work, we selected up to top 10\% behavioral features using 16 feature selection methods (e.g., statistical-based, information theory-based) for the multi-class suicide risk assessment -- high, medium, low. For the cross-validation method, we used 10-folds with two runs to measure both Jaccard Index (JI) and Between Threshold Stability (BTS), which were then used to validate and select the final feature set, as described in \cite{9253541}. Notably, we treated the low, medium, and high categories as nominal instead of ordinal because we are more interested in categorical differences among the three clusters, without assuming that measurements from medium-risk participants always sit between those from low and high-risk counterparts. As such, we chose not to execute correlations analysis where both variables are ordinal or continuous.

\subsection{Statistical Analysis}
A one-way analysis of variance (ANOVA) test was performed for each functional to determine whether a systematic difference between the three suicide risk classes (low, medium, high) exists, followed by a post-hoc two-tailed t-test when statistical significance was found. For both tests, we chose a significance value threshold of 0.05 as alpha, and apply Bonferroni correction to obtain a corrected significance level of $0.05/210 = 2.3809\cdot 10^{-4}$. We selected the functionals for which at least one of the three post-hoc p-values is within this corrected significance threshold. Additionally, we conducted a multivariate analysis of variance (MANOVA) and a two-factor ANOVA to understand subject dependence and selected for the most discriminating feature of suicide risk. Specifically, since the between-subject groups are unbalanced and there is an interaction between independent variables, type-III sums of squares estimation was utilized for the two-factor test. As more than one sample was taken from each video sample the due to the 2-minute segmentation, we ran a repeated-measure ANOVA to determine features that are indicative of suicide risk, accounting for subject-dependence. Although assumptions for ANOVA and T-tests are not fully satisfied -- normality and non-colinearity -- as our sample size is enlarged with video segmentation, we proceed executing the tests with caution.

\subsection{Classification}\label{SCM} 
Our goal is to classify the statistical and selected features into three classes for suicide risk: low-risk, medium-risk, high-risk. To select the best model for this task, we tried variations on three discriminative models: a multilayer perceptron (MLP) with one hidden layer and SVMs with linear and radial kernels. To reduce potential bias caused by overrepresented classes, we tested two different sampling methods: oversampling and undersampling. For oversampling, we randomly duplicated samples from underpopulated classes, and for undersampling, we randomly removed samples from overrepresented classes. This is to ensure that all classes have the same number of samples for training.

For each combination of model and sampling method, we performed 10 random stratified trials using data from all 144 statistical summaries. In each trial, the data was randomly split into a stratified training set and testing set, with a split of around 75\% training data and 25\% test data. Both sets were normalized based on the training data and scaled to values between 0 and 1. The training set was then either left alone, balanced via oversampling, or balanced via undersampling before fitting the model. 

Hyperparameter tuning was done via a grid search using 10-split cross-validation on the training data. For MLP, our only hyperparameter was the number of hidden units. We searched over an evenly spaced range of 10 values slightly greater than the number of features to ensure that a good spread of reasonable hypothesis classes were tested. For testing with 210 features, this range was from 22 to 232 hidden units, and for the selected 12 features, the range was 2 to 14. For SVM, we chose our $C$ value by iteratively performing narrowing grid searches over intervals of values between 0 and 1, eventually narrowing down to values 0.01 apart, then selecting the best one. For SVM with radial kernels, we used several values for gamma, calculated using 1/(n\_features * variance in training sample).

The final performance of each modeling process is represented by its performance on the test data, measured by the mean balanced accuracy score across all 10 trials. As false negatives are highly undesirable when classifying suicide risk, we took measures to ensure classification quality between classes. First, we selected hyperparameters based on balanced scores, which account for the number of samples in each class, and calculated a Matthews Correlation Coefficient (MCC) score for each trial. Accounting for the differences between class sample sizes, MCC is regarded as one of the best evaluators of classification quality in comparison with other measures like F1 score, since it is only high when good results are achieved from all classes \cite{chicco_jurman_mcc}. We used a generalized multiclass version of MCC as described by Gorodkin \cite{gorodkin_mcc}.

We performed two additional randomized experiments. One experiment was done using randomly shuffled feature vectors and the original labels, and another with randomly shuffled labels and the original feature data. By comparing our classification results to these results obtained by chance, we ensure that our results are robust and meaningful. We report all of our results with the best-performing combinations of models and sampling methods.

\section{Results and Discussion}
\subsection{Interpretation of Nonverbal Behavior} 
\subsubsection{Statistical Analysis}

\begin{table}[t]
\caption{Interpretations of the Features that Passed Corrected p-values by Statistical Analysis: 69 behavioral features passed ANOVA and post-hoc t-tests.
}
\begin{center}
\scriptsize
\begin{tabular}{@{}m{8em} | m{8.5cm}| m{1.3cm}@{}}
\hline
\textbf{Behavioral Theme} & \textbf{Statistical Features} & \textbf{Direction} \\ \hline

\multirow{6}{8em}{Deminished Eye and Head movement} & var. of speed of EAR changes, var. of EAR, max. of speed of eye pitch movement, mean of speed of eye pitch movement, rang. of speed of eye pitch movement, std. of speed of eye pitch movement, var. of speed of eye pitch movement, max. of acc. of eye pitch movement, rang. of acc. of eye pitch movement, std. of acc. of eye pitch movement, var of acc. of eye pitch movement, std. of head pitch, var. of head pitch & L \textgreater M \textgreater H \\ \cline{2-3} 
 & std. of speed of EAR changes, std of acc. of EAR changes, max. of EAR, range of EAR, std. of EAR, mean of eye yaw & L \textgreater H \textgreater M \\ \cline{2-3} 
 & rang. of head roll, std. of head roll, rang. of head yaw, std. of head yaw, var. of head yaw & M \textgreater L \textgreater H \\ \cline{2-3} 
 & var. of head roll & M \textgreater H \textgreater L \\ \cline{2-3} 
 & kurt. of speed of head pitch movement, skew of speed of head pitch movement, skew of head roll movement, kurt. of head roll movement & H \textgreater L \textgreater M \\ \cline{2-3} 
 & min. of eye pitch movement, min. of acc. of eye pitch movement, kurt. of speed of head distance movement, skew of speed of head distance movement & H \textgreater M \textgreater L \\ \hline

\multirow{4}{8em}{Anxiety-related involuntary behaviors} & skew of speed of eye yaw movement & L \textgreater M \textgreater H \\ \cline{2-3} 
 & kurt. of head distance & M \textgreater L \textgreater H \\ \cline{2-3} 
 & skew of EAR changes & M \textgreater H \textgreater L \\ \cline{2-3} 
 & peaks of head roll, valys. of head roll & H \textgreater L \textgreater M \\ \hline

\multirow{5}{8em}{Engagment in conversation (head movement to signify emotion, eye contact, head distance)} & kurt. of eye yaw, max. of eye pitch, mean of eye pitch, rang. of eye pitch, std. of eye pitch, var. of eye pitch, kurt. of speed of eye yaw movement, kurt. of acc. of eye yaw movement & L \textgreater M \textgreater H \\ \cline{2-3} 
 & mean of head pitch & L \textgreater H \textgreater M \\ \cline{2-3} 
 & rang. of eye yaw, std. of eye yaw, var. of eye yaw, std. of speed of eye yaw movement, var. of speed of eye yaw movement & M \textgreater H \textgreater L \\ \cline{2-3} 
 & min. of head pitch & H \textgreater L \textgreater M \\ \cline{2-3} 
 & peaks of head pitch, valys. of head pitch & H \textgreater M \textgreater L \\ \hline

\multirow{2}{8em}{Miscellaneous} & min. of eye yaw,  peaks of eye yaw, valys. of eye yaw & L \textgreater H \textgreater M \\ \cline{2-3} 
 & min. of head roll, head\_roll-d1\_kurt & H \textgreater L \textgreater M \\ \hline
 
 \multicolumn{3}{@{}p{28em}@{}}{\footnotesize{\textbf{var.}: variance, \textbf{kurt.}: kurtosis, \textbf{valys.}: number of valleys, \textbf{acc.}: acceleration}}
\end{tabular}
\label{tab5}
\end{center}
\end{table}

ANOVA tests were performed for the 210 functionals collected over two-minute windows, among which, 123 were significant. To further verify the significance, we conducted post-hoc t-tests for each behavioral feature between the three groups.
From this process, 69 functionals were found to be significant in at least one of the three intergroup comparisons (low/high, med/high, low/med). 
Moreover, we used the t-values to determine the direction of the correlations between the groups.

T-tests show that higher risk participants exhibit lower levels of ocular and head activity, as well as heightened anxiety and depression-related symptoms, including ``fight-or-flight'' movements and social withdrawal. Table \ref{tab5} enumerates all 69 features, from which the most prominent ones are discussed in this section.
Higher risk individuals blink less often and have narrower eye-openings and less sudden movements of the eyelids. The minimum of the EAR is zero (when the eye is closed). As low-risk individuals have the highest maximum, range, and standard deviation in EAR, their ocular activity is more diverse. They have wider eye openings and more frequent blink. Additionally, given that the speed and acceleration of the EAR is near-zero when the eyes are open, low-risk participants’ high standard deviations in their EAR's first and second derivatives also signify they blink more and faster. Assuming all interviewees have similar physical needs of blinking for lubricating the eye, the reduction in eyelid activity among high-risk individuals allude to their less robust nervous and muscular system around the eyes.

Low-risk subjects have higher levels of gaze movement, characterized by faster and more frequent up-down glancing. Among all three groups, low-risk individuals have the greatest mean, range, and standard deviation in their eye pitch movement speeds, highlighting they have the most significant gaze shift. The same statistics show that high-risk participant’s up-down gaze shifts have much smaller magnitude. Similarly, low-risk individuals also have more jerky movements, as evidenced by high max., range, std., and var. in the acceleration of eye pitch.
Analysis of head motions yields similar findings---high-risk subjects have the least range of motion when moving their heads in all directions. These trends are inferred from the low range and std. of head roll and yaw among the high-risk participants, and are in line with literature: the slower, diminished, and more sporadic movement of higher risk individuals exhibit signs of fatigue and lethargy that can be attributed to psycho-motor retardation --- one of the most prominent identifiers of depression \cite{fossi1984ethological,6738869,waxer1977nonverbal,alghowinem2013head}.

Statistical analysis also reveals that suicidal individuals have higher levels of anxiety. Notably, the low skew on the speed of high-risk subjects’ eye yaw indicates that they have more frequent left-right movement in their eyes, as they spend more time move their eyes at higher speeds. This corroborates with findings in \cite{MOUKHEIBER2010147,doctorsinterview,HORLEY200333,JUN2013193} that high risk subjects scan the room from left to right and rarely make eye contact with the interviewer. Moreover, despite having the smallest range in motion, high-risk group tilt, nod, and move their head more than their lower-risk counterparts, as signified by the low range, var. and std. of head roll and yaw.

This study observed augmented levels of social withdrawal among higher risk individuals. Statistics pertaining to the left-right movement of the gaze indicates that low-risk participants’ gaze is often focused on the interviewer---they rarely look away, and even when they do so, their gaze would quickly come back to the interviewer. Additionally, low-risk subjects’ gaze and head pose direction are the most elevated. These findings corroborate with the correlation between depression, anxiety, and eye contact avoidance found by previous studies \cite{MOUKHEIBER2010147,HORLEY200333,JUN2013193,6738869,waxer1977nonverbal,alghowinem2013head}.

Medium-risk participants' features frequently sit between low and high-risk groups in most behaviors. The level of their eyelid and gaze activity is higher than high-risk individuals, but lower than low-risk individuals; their vertical gaze direction is higher than high-risk, but lower than low-risk subjects. These trends affirm that medium-risk participants experience moderate drowsiness in movement and depression comparing to high and low-risk participants. Other trends pertaining to medium-risk participants, however, call for further investigation.They have the least front-back head movement, as they seldom lean forward or back to show engagement in the conversation. They blink the slowest, although the difference is not significant between high and medium-risk groups. They tilt their head the least often compare to low and high-risk group, which can be interpreted as either high level of focus or severe psycho-motor retardation. When scanning the room, medium-risk subjects' gaze diverges the furthest away from the interviewer, although their left-right gaze movement is sparse. This may be interpreted as fixation at a spot away from the interviewer, either due to curious distraction or eye contact avoidance. While these trends are insightful, additional information is required for further interpretation.

\begin{table}[t]
\caption{Interpretations of the 21 Behavioral Features Selected that Passed the 2-Factor ANOVA 
}
\begin{center}
\scriptsize
\begin{tabular}{@{}m{4.5cm} | m{4.7cm}| c@{}}
 \hline
\textbf{Behavioral Theme} & \textbf{Functional Features} & \textbf{Direction} \\ \hline
\multirow{4}{3.5cm}{Engagement in conversation\\ (head movement to signify emotion, eye contact, head distance)} & range of head pitch & L \textgreater M \textgreater H \\
 
 & skew of head distance & M \textgreater L \textgreater H \\
 
 & std. of head distance & M \textgreater H \textgreater L \\
 & skew of EAR &  H \textgreater L \textgreater M \\
 \hline
 
\multirow{4}{10em}{Room scanning (anxiety symptom)} & kurt. of acc. of eye pitch & L \textgreater  H \textgreater M \\

 & min. of speed of head yaw & H \textgreater L \textgreater M \\
 & kurt. of head roll & \\
 & min. of speed of head yaw & \\
 \hline

\multirow{4}{10em}{Diminished Eye and Head movement (depression symptom)} &skew of speed of eye pitch & L \textgreater  H \textgreater M \\

 & range of head pitch & L \textgreater M \textgreater H \\
 & var. of speed of head pitch & \\
 & var. of acc. of head pitch & \\
 & var. of acc. of head roll & \\
 & var. of acc. of eye pitch & \\
 
 & min. of speed of EAR & M \textgreater H \textgreater L \\
 
 & skew of EAR & H \textgreater L \textgreater M \\
 & kurt. of head yaw & \\
 & min. of speed of head yaw & \\
\hline
 
\multirow{3}{10em}{Miscellaneous} & kurt. of speed of EAR & M \textgreater L \textgreater H \\

 & skew of head yaw & H \textgreater L \textgreater M \\
 & skew of head roll & \\
 \hline
\multicolumn{3}{l}{\footnotesize{\textbf{var.}: variance, \textbf{valys.}: number of valleys, \textbf{acc.}: acceleration}}

\end{tabular}
\label{tab10}
\end{center}
\end{table}

Multivariate Analysis of Variance (MANOVA) determines no significant difference in the extracted features for the three suicide levels. Roy’s Greatest Root is the only statistic that yields a significant p-value \footnote{Detailed MANOVA results are presented here: https://bit.ly/32h6CRa}. This is expected, as the feature space is especially large. 
Using the repeated-measure ANOVA, we found suicide risk to have a significant effect in the number of sudden movements in participants' left-right eye gaze. T-tests subsequently found the low-risk group have the highest number of valys. of eye yaw acc., hence the most sudden movements, while high-risk group to have the least. This finding corroborates the depression-induced psycho-motor retardation hypothesis.

Two-factor ANOVAs were conducted for each feature in order to further understand subject and risk-level-based dependence. Specifically, we conducted type-III ANOVA, which assumes dependence between independent variables -- risk level and identity of the participant --  which are not independent. For all features, we compute the probability of the distribution being identical for both variables and their interaction, and selected 18 features that are both significant in level-identity interaction and insignificant for identity differences (alpha $< 0.05$). As such, we deem the 18 features to be free of subject-dependence. Among the 18 features, one was selected by features selection, and one by t-tests.

Analysis of the 18 features using the group-wise t-statistic found remarkable trends. In general, participants at higher suicide risk exhibit more distinguishable depression symptoms, and those at lower risk, the least. Specifically, since  head pitch's velocity, var. of acc., and range are high, we deem the low-risk group to have the most frequent sudden up-down head movements, the greatest range of head-nodding motion, and move their heads the quickest when they do so. The low-risk group also have the most sudden head tilts and fastest sudden up-down movements in their eye gaze, as signified by the variance in acceleration in head roll and max. acc. in eye pitch. Participants in the high-risk group display the least of these eye and head activities. Additionally, high-risk participants rarely turn or tilt their heads (high kurt. in head yaw and roll), and have low levels of eye openness (high skew in EAR). Such ocular and head movement reduction can be explained with depression-induced psycho-motor retardation and social disengagement. In contrast, the low-risk group often lean forward as they engage in conversation with the interviewer, as observed by the high std. of their head-distance. This signal, in combination with low-risk participants' higher activity level, signify their higher level of social engagement.

High and medium-risk groups also exhibit anxiety symptoms. Especially, medium-risk individuals have the quickest and most frequent head turns and tilts, given the low skew in head yaw and roll. As the interviewees' gaze is in the direction of the interview most of the time, the medium-risk group's least uniform velocity in their eyes' up-down movements signify that they frequently look away, up-down scanning the room.
 
 Some findings of 2-factor ANOVA lack explanation. Medium-risk group have the highest kurt. of speed of EAR, meaning that the speed in their eyelid's movement is the most uniform. This implies they blink the least often. It was also found that medium-risk group have the lowest skew in head yaw and roll, and high-risk group have the highest. This implies that medium risk group tilt their heads to the right more often while the high risk group, more to the left. This may be due to fixation at a spot away from the interviewer, yet further investigation is required for a more meaningful interpretation.

\subsubsection{Feature Selection}

\begin{table}[t]
\caption{Interpretations of the Features Selected by Feature Selection Framework: 12 behavioral features were narrowed down through the framework
}
\begin{center}
\scriptsize
\begin{tabular}{@{}m{15em} | m{5.7cm}| m{1.5cm}@{}}
 \hline
\textbf{Behavioral Theme} & \textbf{Functional Features} & \textbf{Direction} \\ \hline
\multirow{7}{10em}{Engagement in conversation\\ (head movement to signify emotion, eye contact, head distance)} & std. of head pitch & L \textgreater M \textgreater H \\
 & avg. of eye gaze pitch & \\
 
 & std. of head yaw & M \textgreater L \textgreater H \\
 & var. of head yaw & \\
 
 & min. of head distance & M \textgreater H \textgreater L \\
 & std. of eye gaze yaw & \\
 & var. of eye gaze yaw & \\
 \hline
 
Room Scanning (anxiety related) & var. of the speed of eye gaze yaw shift & M \textgreater H \textgreater L \\ \hline

\multirow{4}{10em}{Diminished Eye and Head movement} & std. of the speed of EAR change & L \textgreater  H \textgreater L \\

 & valys of acc. of eye gaze yaw shifts & L \textgreater M \textgreater H \\
 & var. of acc. of head pitch change & \\
 
 & skewness of acc. of head roll change & M \textgreater H \textgreater L \\
 
 \hline
\multicolumn{3}{l}{\footnotesize{\textbf{var.}: variance, \textbf{valys.}: number of valleys, \textbf{acc.}: acceleration}}

\end{tabular}
\label{tab3}
\end{center}
\end{table}

\begin{table*}[t]
\caption{The Best Classification Results using Statistical Summaries Obtained from 2-Minute Windows}
\centering
\scriptsize
\begin{tabular}{l||m{3.8em}|l|m{3.5em}||l|m{3.8em}|m{3.8em}||m{4.0em}||m{4.5em}}
\hline
 \textbf{Features}                    & \multicolumn{3}{c||}{\textbf{All features}}                                                    & \multicolumn{3}{c||}{\textbf{Selected features}}                                               & \textbf{Random features}       & \textbf{Shuffled labels}                \\
                     \hline
 \textbf{Performance}                    & \multicolumn{1}{c|}{Best} & \multicolumn{1}{c|}{2nd} & \multicolumn{1}{c||}{3rd} & \multicolumn{1}{c|}{Best} & \multicolumn{1}{c|}{2nd} & \multicolumn{1}{c||}{3rd} & \multicolumn{1}{c||}{Best} & \multicolumn{1}{c}{Best} \\
  \textbf{}                    & \multicolumn{1}{c|}{Model} & \multicolumn{1}{c|}{best} & \multicolumn{1}{c||}{best} & \multicolumn{1}{c|}{Model} & \multicolumn{1}{c|}{best} & \multicolumn{1}{c||}{best} & \multicolumn{1}{c||}{Model} & \multicolumn{1}{c}{Model} \\
                     \hline
sampling method      & none                           & none                         & over-sample                   & none                           & none                         & over-sample                   & under-sample                    & over-sample                     \\
model                & SVM (linear)                   & MLP                          & SVM (linear)                 & MLP                            & SVM (radial)                 & SVM (linear)                 & SVM (radial)                   & MLP                            \\
avg accuracy         & 0.983                          & 0.978                        & 0.977                        & 0.977                          & 0.971                        & 0.961                        & 0.425                          & 0.399                          \\
std. accuracy & 0.019                          & 0.020                        & 0.033                        & 0.021                          & 0.023                        & 0.033                        & 0.045                          & 0.077                          \\
avg MCC              & 0.966                          & 0.970                        & 0.957                        & 0.966                          & 0.974                        & 0.943                        & 0.138                          & 0.106                         \\
\hline
\end{tabular}

\label{tab:classification_results}
\end{table*}

A feature selection method developed by \cite{9253541} was applied on our suicide data set. Using the same 210 functionals' statistical summaries as inputs, a total of 12 features passed the FS framework.
To investigate the direction of these features in each risk group, the state value of the t-test analysis was obtained as listed in Table \ref{tab3}. 
The results from FS align with those obtained from statistical analysis. We observed that, in comparison to the low and medium-risk participants, the high-risk group rarely nod and shake their heads. Their heads' range of up-down movements are the smallest; their sudden left-right movement of the gaze is the least frequent; their eyes' blinking velocity is the most uniform --- indicating less and slower blinks. These observations lead us to conclude that, due to depression-induced psycho-motor retardation, higher-risk individuals exhibit reduced activity in their heads and eye regions \cite{fossi1984ethological}.

Results from feature selection also shed light on the participants' emotional expression. The low mean in the eye-pitch and the high mean in the eye yaw indicate that medium and high-risk subjects spend significantly more time looking down and avoiding eye contact with the interviewer --- strong signals of social withdrawal \cite{waxer1977nonverbal,fossi1984ethological}. The low-risk group spent more time tilting their heads and leaning forwards as they engage with the interviewer. The eye's left-right moving speed is the most uniform among low-risk subjects, which parallels previous works --- anxiety patients often have a hard time focusing on the interviewer and scan the room. Given the prevalence of depression and anxiety among suicidal patients, these results are in line with literature, that depression leads to psycho-motor retardation and anxiety gives rise to more frequent distraction from the conversation.

\subsection{Classification Results}

Table \ref{tab:classification_results} shows the classification results using the best-performing classifiers, measured by balanced accuracy\footnote{Full classification results are presented here: https://bit.ly/3tnqcY1}. From our top 3 best-performing classifiers, we can see that all models performed similarly well. Still, the best model accuracy-wise for all features was an SVM with linear kernels, while the best for selected features was MLP.
Even though the best balanced accuracies were 98.3\% and 97.7\% with full and selected features, respectively, the best MCC scores were obtained from our 2nd best models (0.970 and 0.974 for full and selected features respectively). 
Given that MCC accounts for the imbalanced samples from each class, we can argue that our 2nd best models performed better than the best (even though with small margin) since it has less confusion between the classes.
For both all features and selected features, training using the original dataset worked far better than undersampling (4.5\% absolute difference across all trials) and slightly better than oversampling (0.6\% absolute difference).

Using all statistical features for training, our best model averaged 98\% accuracy with a small standard deviation. The mean MCC was 0.967 --- almost perfect test performance across all three classes. We achieved similar numbers using the selected features with slightly greater standard deviation. Since we were able to immensely reduce the feature space with the FS framework without impacting classification accuracy, we conclude that the 12 selected behavioral features can indicate suicide risk level as well as represent the model for a behavioral interpretation.

For comparison, running the same tests using randomly switched labels resulted in far worse accuracy (ranging from 30-40\%), as did the random feature vectors (also ranging from around 30-40\%). Thus, we are confident that our results are indicative of the discriminative power of our features, not of any algorithm bias. In addition, achieving high accuracy with both MLP and SVM implies that our features are easily separable between classes, so they would hold good discriminative power regardless of the modeling method.
Overall, our results indicate that both the entire statistical summary and the selected features are excellent at distinguishing between different levels of risk with high accuracy.


\subsection{Limitations}
We acknowledge the small sample size of the dataset, and our results necessitates further investigation upon features that cannot be explained. Particularly, high-risk participants have the greatest head tilt to their left, but the speed of their head tilt movement is the most uniform. Moreover, while our results support the psycho-motor retardation theory, we found the high-risk group to tilt and nod their head frequently but with the smallest range of motion. We hypothesized that such contradicting results stem from the association between suicide risk and bipolar disorder \cite{jamison2000suicide,plans2019association}. Since bipolar disorder patients experience both high and low energy episodes --- which could sufficiently interpret our findings --- we hope to also explore the biomarkers of bipolar disorder.
Additionally, studies on eye gaze of people suffering from mental disorders shows nonspecific gaze \cite{schelde1998major}, anxious gaze aversion \cite{perez2003nonverbal}, or simply gazing at the door during uncomfortable situations \cite{Grandin1995,Eisenberg2004Emotion}.
As the interviews were conducted in different locations, adding the room context to the feature set will provide insight into what the subject is gazing at.

\section{Conclusion}

Suicide poses significant risk, to those at risk, their families, and our society. Suicide is also often connected with depression, anxiety, and self-harm, but none of these factors alone is deterministic enough for an objective and effective suicide risk assessment. Additionally, due to privacy constraints, biomarkers of suicide have received very limited attention in academia and industry, as datasets that contain sensitive information are rarely  available.

In this work, we aim to extend and fill the gap by systematically and objectively exploring nonverbal behaviors. We employ statistical analysis, feature selection, and machine learning classification to analyze signals from the eyes and head. Applying these three methods to our dataset, our conclusion corroborates previous psychological studies. Particularly, high-risk suicidal subjects exhibit lower activity levels in their eyes and head movements and are subject to a higher degree of psycho-motor retardation. Their eye gaze shows symptoms of anxiety and depression, including constant left-right scanning, difficulty of maintaining eye contact, and angling down their gaze and head pose. 

Our classification results showed that these features are representative of suicidal intent. We achieved classification accuracy ranging from 96-98\% for three risk levels across different modeling methods. We also highlight that since the results from statistical analysis and feature selection lead to the same conclusion, both are viable means of objective suicide risk assessment.
Yet, conclusions from this work must be tested and verified against a larger and more diverse dataset. Future work will also extend to study body movements and speech prosody as potential biomarkers of suicide intents. 

\section*{Acknowledgement}
We thank and acknowledge the effort made by NHK, Nippon Hoso Kyokai (Japan Broadcasting Corporation)  for conducting, recording and providing the interview dataset used in this work.

\bibliographystyle{splncs04}
\bibliography{ref}

\end{document}